\documentclass{IEEEtran}
\usepackage{spconf,amsmath,graphicx}

\usepackage[dvipsnames]{xcolor}
\usepackage{booktabs}

\usepackage[noadjust]{cite}
\usepackage{balance}

\usepackage{algorithm}

\usepackage{lipsum}

\usepackage[binary-units=true]{siunitx}

\usepackage[noend]{algpseudocode}  % Should be loaded after hyperref
\algrenewcommand\algorithmicindent{0.5em}%

\usepackage{setspace}

\usepackage{subcaption}

\usepackage{tikz}
\usepackage{pgfplots}
\pgfplotsset{compat=1.10}   %% <-- this added

\definecolor{vlightred}{RGB}{217,150,144}
\definecolor{vheavyred}{RGB}{147,58,50}

\definecolor{vlightgreen}{RGB}{216,234,180}
\definecolor{vheavygreen}{RGB}{121,149,64}

\definecolor{vlightblue}{RGB}{167,225,241}
\definecolor{vheavyblue}{RGB}{49,133,156}

\definecolor{vlightyellow}{RGB}{255,217,102}
\definecolor{vheavyyellow}{RGB}{191,144,0}

\newcommand{\tikzcircle}[2][red,fill=red]{\tikz[baseline=-0.5ex]\draw[#1,radius=#2] (0,0) circle ;}%

\makeatletter
\let\OldStatex\Statex
\renewcommand{\Statex}[1][3]{%
	\setlength\@tempdima{\algorithmicindent}%
	\OldStatex\hskip\dimexpr#1\@tempdima\relax}
\makeatother

\title{Lupulus: A Flexible Hardware Accelerator for Neural Networks}
\name{\vspace{-5mm}Andreas Toftegaard Kristensen\IEEEauthorrefmark{1}, Robert Giterman\IEEEauthorrefmark{1}, Alexios Balatsoukas-Stimming\IEEEauthorrefmark{2}, and Andreas Burg\IEEEauthorrefmark{1}}
\address{\IEEEauthorrefmark{1}Telecommunication Circuits Laboratory, \'{E}cole polytechnique f\'{e}d\'{e}rale de Lausanne, Switzerland\\
\IEEEauthorrefmark{2}Department of Electrical Engineering, Eindhoven University of Technology, Netherlands}

\begin{document}

\maketitle
\begin{abstract}
Neural networks have become indispensable for a wide range of applications, but they suffer from high computational- and memory-requirements, requiring optimizations from the algorithmic description of the network to the hardware implementation.
Moreover, the high rate of innovation in machine learning makes it important that hardware implementations provide a high level of programmability to support current and future requirements of neural networks.
In this work, we present a flexible hardware accelerator for neural networks, called \emph{Lupulus}, supporting various methods for scheduling and mapping of operations onto the accelerator.
Lupulus was implemented in a \SI{28}{\nano\metre} FD-SOI technology and demonstrates a peak performance of $380\,\mathrm{GOPS/GHz}$ with latencies of \SI{21.4}{\milli\second} and \SI{183.6}{\milli\second} for the convolutional layers of AlexNet and \mbox{VGG-16}, respectively.
\end{abstract}

\section{Introduction}
\label{sec:intro}

%%%%% Opening (teaser) + identify the problem %%%%%

\emph{Neural networks} (NNs) have state-of-the-art performance for a wide range of applications, including speech-recognition~\cite{hinton2012deep}, time-series forecasting~\cite{schmidhuber2015deep}, and computer vision~\cite{LeCun2015}.
However, this performance comes at the cost of high computational complexity and storage, as billions of multiply-accumulate (MAC) operations and many megabytes of memory for the NN parameters are required~\cite{He2015}.
This is a problem in resource- and energy-constrained devices, which necessitates optimizations from the algorithmic description of the NN down to the hardware.

%%%%%% Why do we want to do what we do %%%%%
%
Several hardware accelerators (HwAs) have recently been proposed to optimize the execution of NNs~\cite{Chen2016, han2016eie, wei2017automated, du2017reconfigurable, andri2018yodann, hegde2018ucnn, yin2017high, desoli2017, luo2016dadiannao}.
These HwAs achieve high performance by exploiting the inherent parallelism of NNs, splitting the computations across hundreds of processing elements (PEs) with maximum data-reuse within and across PEs using small local memories.
While these HwAs are relatively simple at a high-level, the large number of design parameters and requirements for different NNs lead to complex design decisions.
Moreover, given the high cost and effort of FPGA/ASIC implementations and the rate of innovation in machine learning, HwAs should provide a high level of programmability to support the current and future requirements for NNs, while maintaining a high utilization of the hardware resources.

\emph{Contribution:} In this paper, we describe \emph{Lupulus}, a flexible HwA for NNs, which supports a variety of NN architectures by applying different scheduling and operation mapping strategies.
We synthesize Lupulus using a \SI{28}{\nano\metre} FD-SOI technology with a \SI{1}{\volt} operating voltage and a target frequency of \SI{1}{\giga\hertz}, providing a theoretical peak performance of $380\,\mathrm{GOPS}$.
Results for NN execution time show that Lupulus is capable of efficiently executing the different layers of AlexNet and \mbox{VGG-16}, and outperform a similar accelerator on \mbox{VGG-16} when on-chip resources and memory interface bandwidth are matched.

\emph{Relation to Previous Work:}
Many existing accelerators, such as~\cite{Chen2016, han2016eie, du2017reconfigurable, wei2017automated} have small local memories in the PEs for the weights, inputs, and partial sums, which may leave the local memories for the partial sums underutilized when partial sums are forwarded to a neighboring PE instead of being stored in the PE itself.
Moreover, the partial sums may have to be read out to a high-level memory and then sent back later if the local memories are too small.
In our case, the partial sums are stored for groups of PEs, making it easier to fully utilize the memory for the partial sums.
A similar architecture to Lupulus is~\cite{du2017reconfigurable}, which also uses $3\times 3$ blocks of PEs.
However, for $1 \times 1$ convolutions, only two PEs out of nine can be turned on in~\cite{du2017reconfigurable}, whereas our architecture can use all PEs.
\looseness=-1

\section{Background}
\label{sec:background}

In a NN, the raw input data is transformed into high-level abstract representations to extract useful information in a process called \emph{inference}, involving multiple stages of non-linear processing, each of which is referred to as a layer. 
In \emph{feed-forward neural networks} (FFNNs), the execution of a layer is given as
\begin{equation}\label{eq:inference}
\mathbf{a}^{[L]} = f( \mathbf{W}^{[L]} \mathbf{a}^{[L-1]} + \mathbf{b}^{[L]} ) \, ,
\end{equation}
where $\mathbf{a}^{[L-1]}$ is the output from layer $L-1$, $\mathbf{W}^{[L]}$ are the network weights, $\mathbf{b}^{[L]}$ are the biases in layer $L$, and $f(\cdot)$ is a non-linear \emph{activation function}. 
\emph{Convolutional neural networks} (CNNs) are the most common deep neural networks for image processing. 
For a CNN, the matrix-vector product in~\eqref{eq:inference} is replaced by the cross-correlation operation, which for a 2-D input \emph{feature map} $F$ is defined as $G[i,j]=\sum_{u=-k}^{k} \sum_{v=-k}^{k} h[u,v] F[i+u, j+v]$, where $h$ is the \emph{kernel/filter} of shape $(2k+1) \times (2k+1)$.
When CNNs are used with 3-D inputs, such as RGB images, the input feature map consists of multiple 2-D planes, called channels, and the weight matrix $\mathbf{W}$ is composed of multiple 3-D filters/kernels, individually applied to the input feature map~\cite{lecun2010convolutional}.\looseness=-1
%Additionally, a 1-D bias is added to the results for each set of 3-D filters.
%CNNs usually also apply FFNNs as the final layers, denoted as fully-connected (FC) layers.

\begin{algorithm}[t]
	\caption{Cross-correlation algorithm for CNNs, where $\mathbf{I}$ is the input feature map, $\mathbf{W}$ are the weights, and $\mathbf{O}$ is the output feature map. 
	The height, width, and number of channels of a data-structure $\mathbf{A}$ are denoted by $h_{\mathbf{A}}$, $w_{\mathbf{A}}$, and $c_{\mathbf{A}}$, respectively, and $s$ is the stride across $\mathbf{I}$.}
	\label{alg:conv}
	\begin{spacing}{0.93}
		\begin{algorithmic}[1]
			\Require{$\mathbf{I}[c_{\mathbf{I}}][h_{\mathbf{I}}][w_{\mathbf{I}}], \mathbf{W}[c_{\mathbf{O}}][c_{\mathbf{I}}][h_{\mathbf{F}}][w_{\mathbf{F}}], \mathbf{O}[c_{\mathbf{O}}][h_{\mathbf{O}}][w_{\mathbf{O}}]$.}
			\For{$g_{\mathbf{O}} \gets 1$ to $c_{\mathbf{O}}$}
			\State{$\mathbf{F} \gets \mathbf{W}[g_{\mathbf{O}}]$}
			\For{$i_{\mathbf{O}} \gets 1$ to $h_{\mathbf{O}}$}
			\For{$j_{\mathbf{O}} \gets 1$ to $w_{\mathbf{O}}$}
			\For{$g_{\mathbf{I}} \gets 1$ to $c_{\mathbf{I}}$}
			\For{$i_{\mathbf{F}} \gets 1$ to $h_{\mathbf{F}}$}
			\For{$j_{\mathbf{F}} \gets 1$ to $w_{\mathbf{F}}$}			
			\State $\mathbf{O}[g_{\mathbf{O}}][i_{\mathbf{O}}][j_{\mathbf{O}}] \gets \mathbf{O}[g_{\mathbf{O}}][i_{\mathbf{O}}][j_{\mathbf{O}}]$
			\Statex[20] $+$ $\mathbf{I}[g_{\mathbf{I}}][s \times i_{\mathbf{O}} + i_{\mathbf{F}}][s \times j_{\mathbf{O}} + j_{\mathbf{F}}]$
			\Statex[20] $\times$ $\mathbf{F}[g_{\mathbf{I}}][i_{\mathbf{F}} ][j_{\mathbf{F}}]$
			\EndFor
			\EndFor
			\EndFor
			\EndFor		
			\EndFor				
			\EndFor		
		\end{algorithmic}
	\end{spacing}
\end{algorithm}

\subsection{Design Challenges and Optimizations}\label{ssec:challenges}
While NNs consist mainly of MAC operations, modern NNs come in many shapes and sizes~\cite{krizhevsky2012imagenet, simonyan2014very, szegedy2015going}.
These variations in layer size and parameters present a problem for efficiently executing different networks on a custom HwA. 
The full sequence of operations for a CNN can also be represented as multiple nested loops, as shown in Alg.~\ref{alg:conv}.
For a given number of PEs and some amount of local memory, this loop structure can be optimized to better utilize the available resources and reduce the required memory access bandwidth, posing significant challenges on operation scheduling and mapping to maximize data-reuse~\cite{Sze2017}.
\emph{Input reuse} relates to the reuse of the input feature map $\mathbf{I}$.
Given a filter of shape $h_{\mathbf{F}} \times w_{\mathbf{F}}$ and $c_{\mathbf{O}}$ filters, each pixel in the input feature map can be reused approximately $h_{\mathbf{F}} \times w_{\mathbf{F}} \times c_{\mathbf{O}}$ times, depending on the padding and stride.
In \emph{partial sum reuse}, the partial sums from the inner-loop in Alg.~\ref{alg:conv} can be shared between multiple PEs operating in parallel and stored locally until the final result can be delivered.
Furthermore, given an output feature map of shape $h_{\mathbf{O}} \times w_{\mathbf{O}}$, \emph{convolutional reuse} allows each filter to be reused $h_{\mathbf{O}} \times w_{\mathbf{O}}$ times for the same input feature map channel as the filter is slid across it.
Finally, multiple input sets are usually grouped into batches of size $N$, providing \emph{weight reuse} as the same weights can be used for all batches.
Weight reuse is only beneficial for increasing the throughput and experience has shown that most users prefer reduced latency when deploying NNs~\cite{jouppi2017datacenter}.
Therefore, we only consider input, partial sum, and convolutional reuse.
Since different NNs (and even different layers) favor different types of reuse under a set of hardware constraints, the challenge is to provide efficient support for all or most of them in an accelerator.
%\looseness=-1

\begin{figure}[t]
	\centering
	\includegraphics[trim=0.4cm 0.1cm 2.9cm 0.2cm, clip, width=\linewidth]{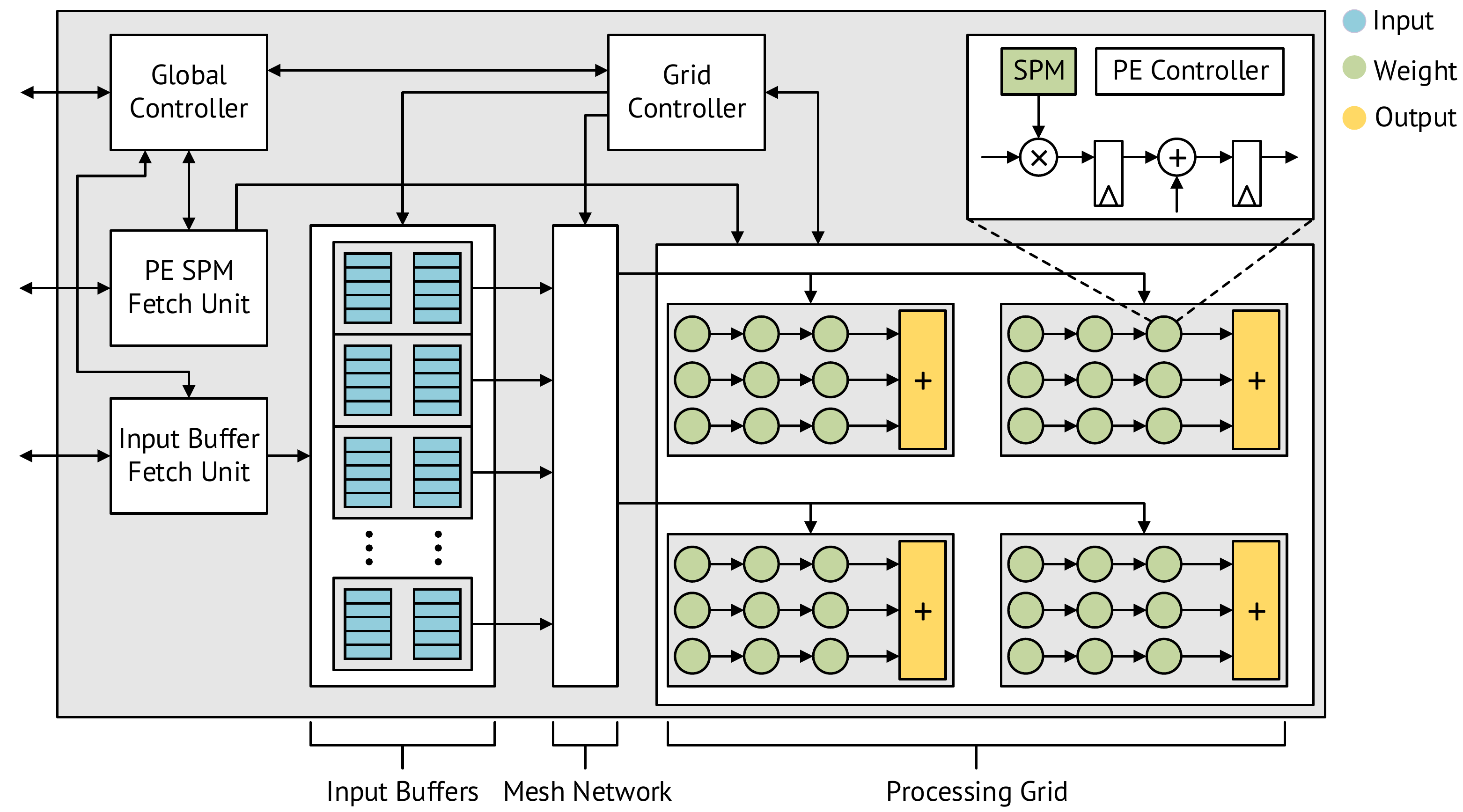}
	\caption{Top-level view of Lupulus, with details of a single PE in the top-right corner. The inputs, weights, and outputs are stored stored in the input buffers (\tikzcircle[vlightblue, fill=vlightblue]{3pt}), in the SPMs of the PEs (\tikzcircle[vlightgreen, fill=vlightgreen]{3pt}), and in the accumulators (\tikzcircle[vlightyellow, fill=vlightyellow]{3pt}), respectively.}
	\label{fig:accelerator}
\end{figure}

\section{Lupulus: Design and Implementation}
\label{sec:design}

%In the previous section, we introduced NNs and the main challenges for implementing NN HwAs.
In this section, we present the design and implementation of our proposed HwA, Lupulus, by providing a full system overview.
Then, we consider the scheduling and mapping strategies supported by the HwA, which enables both data-reuse and executing different NN layers.

\subsection{High-Level Architecture}

A top-level view of Lupulus is shown in Fig.~\ref{fig:accelerator}.
The HwA is composed of controllers, a PE grid, input buffers, and a mesh network connecting the input buffers to the PE grid.

\begin{figure}[t]
	\centering
	\begin{subfigure}{0.95\columnwidth}
		\centering
		\includegraphics[trim=0.2cm 0.1cm 0cm 0.1cm, clip, width=\linewidth]{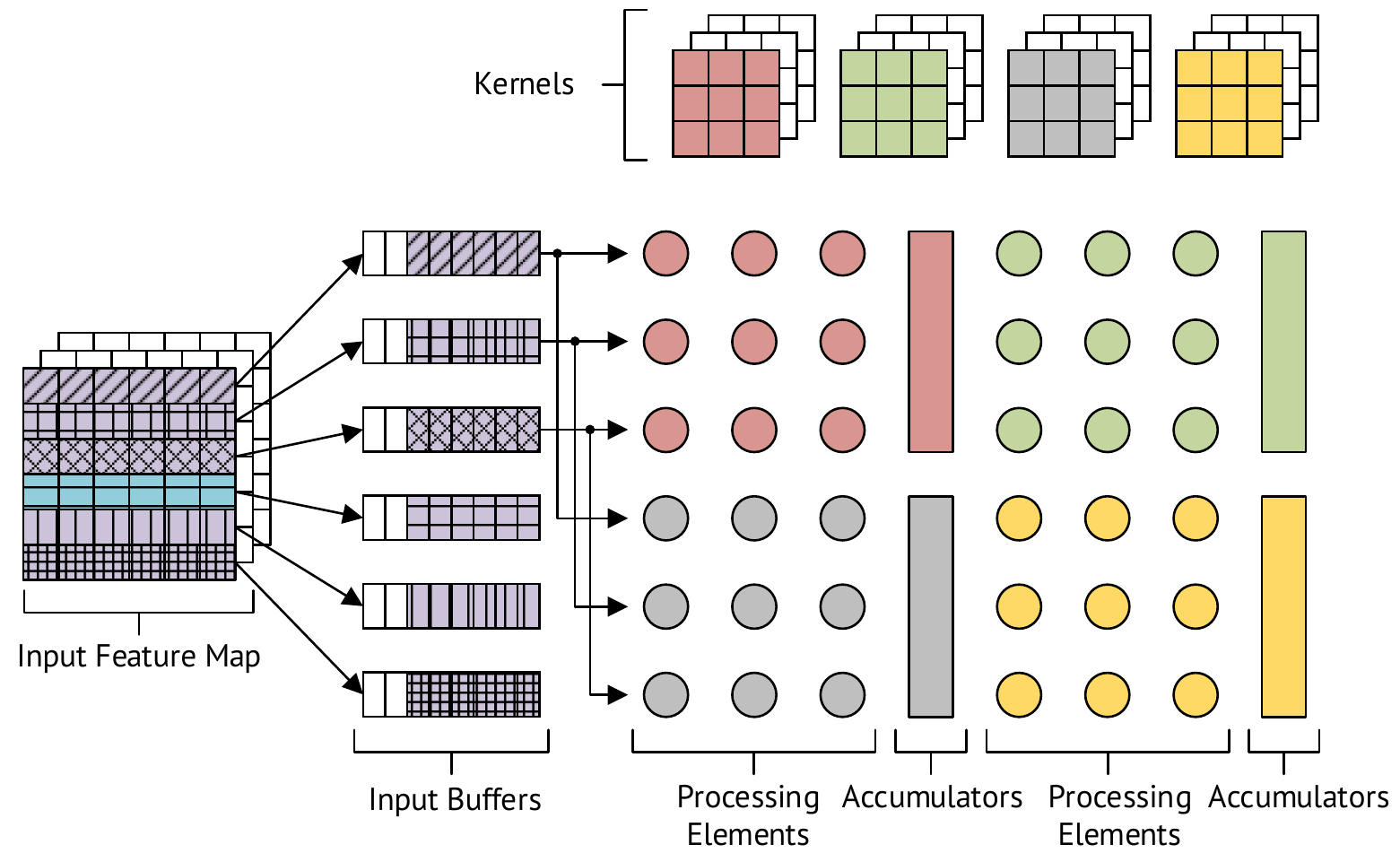}
		\caption{Mapping and scheduling of four $3 \times 3$ kernels and one channel of a $6 \times 6$ input feature map.}
		\label{fig:conv3x3}
	\end{subfigure}
	
	\begin{subfigure}{0.95\columnwidth}
		\centering
		\includegraphics[trim=0.3cm 0.0cm 0cm 0.0cm, clip, width=\linewidth]{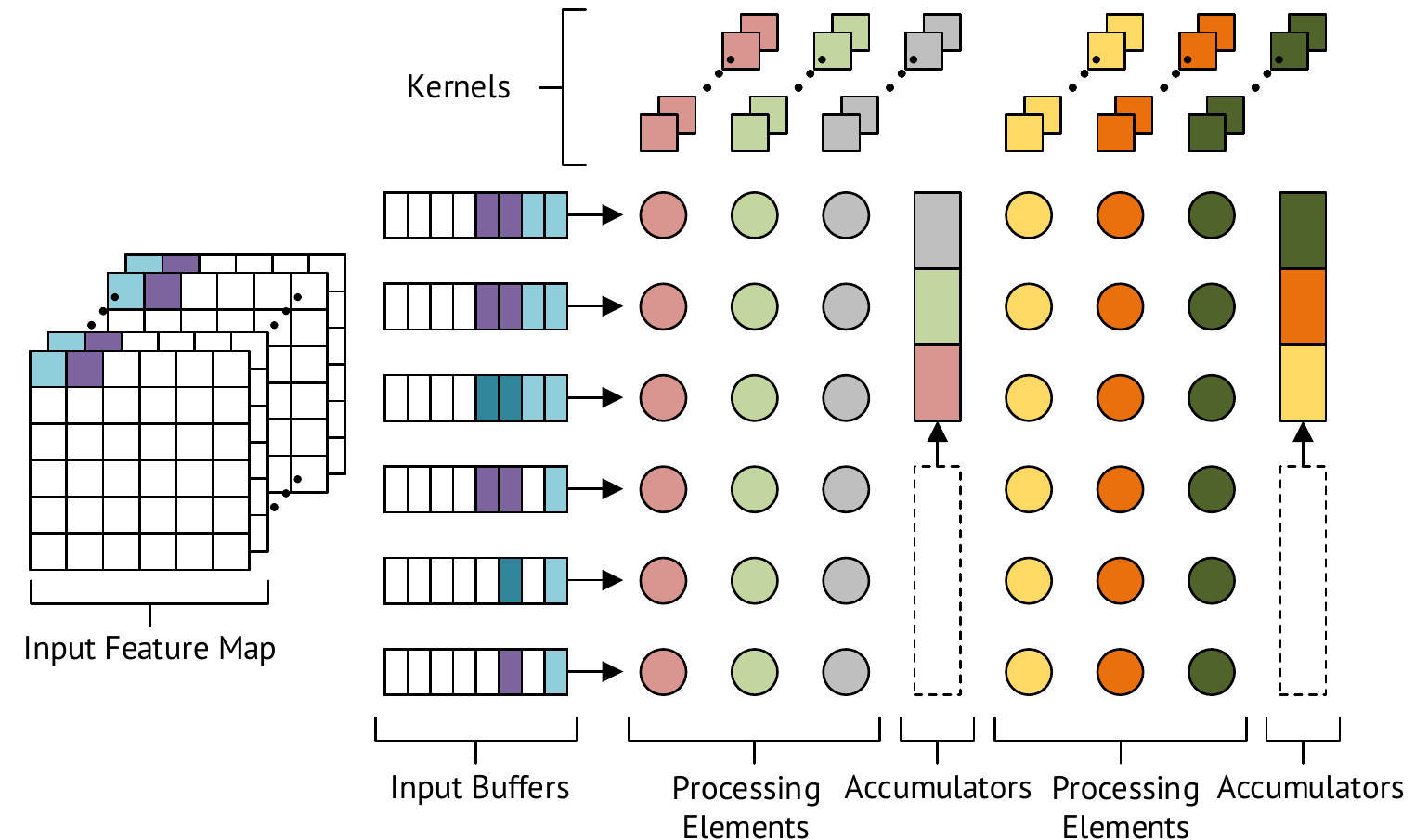}	
		\caption{Mapping and scheduling of $1 \times 1$ kernels and two channels of a $6 \times 6$ input feature map.}
		\label{fig:conv1x1}
	\end{subfigure}%
	\caption{Mapping and scheduling of filters on a $6 \times 6$ PE grid.}
	\label{fig:scheduling}
\end{figure}

The \emph{global controller} is responsible for dispatching offline generated instructions to the fetch units, the processing grid controller and the PEs.
The global controller also controls which fetch units are running, and when the processing grid can start processing the input feature map in the input buffers and the weights in the PEs.
The \emph{fetch units} are programmed with a loop structure, similar to Alg.~\ref{alg:conv}, specifying when to fetch different segments of the weights and inputs and how to map these segments to the HwA.
The \emph{input buffers} contain one SRAM instance per row of PEs.
Extra memory banks for \emph{double-buffering} can be added to enable the overlapping of fetching new inputs and the computation.
The input buffers connect to the PEs through a \emph{mesh network}, which allows for different ways of connecting the memories to the PEs.
Connections can be one-to-one as shown in Fig.~\ref{fig:conv1x1} or one-to-many as shown in Fig.~\ref{fig:conv3x3}.
The \emph{processing grid} contains the PE groups.
Each PE group contains a compile-time configurable number of PEs, which can be optimized for a specific kernel size, and an accumulator module used to accumulate values across the rows of PEs in the group and store the partial sums.
%This means that the size of the PE groups can be optimized for a specific kernel size, while still supporting other shapes.
Small mux networks share the partial sums between the PEs before reaching the accumulators and allow for different groups of PEs to be merged.
The instructions received from the global controller configure the PEs for different padding and striding options.
%The instructions received from the global controller configure the local controller in the PEs, which is used to skip inputs when padding or different stride sizes are used.
%
Finally, the \emph{processing grid controller} configures the mesh network between the input buffers and the PEs based on the NN layer parameters and it indicates to the PEs which weights to use from their local scratch-pad memories (SPMs).
It also communicates the status of the processing back to the global controller to indicate when data can be freed from the input buffers, the SPMs in the PEs, and the accumulators.

\subsection{Operation Scheduling and Mapping}
\label{ssec:scheduling}

We now consider the scheduling and mapping of the data to the HwA, where scheduling refers to when different parts of the data-structures are operated on and the mapping describes on which PE the associated operations are executed.

Fig.~\ref{fig:conv3x3} provides an example with $3 \times 3$ kernels, in this case, the kernel size matches the PE group and the weights of each kernel are mapped directly to a single group of PEs, indicated by the uniformly colored PEs in Fig.~\ref{fig:conv3x3}. 
The inputs can be used for all the filters which can be scheduled simultaneously.
If not all filters fit, the input values are loaded back later, lowering the input reuse.
In cases where the kernel is larger than the PE group, such as a $5 \times 5$ kernel, the kernel is decomposed along the rows into smaller chunks and extended horizontally into the next PE group. 
Kernels can only be decomposed along the rows, meaning that PE grid width is the main limitation in terms of mapping convolutional kernels.
However, as small kernels are the norm today, this is not a limiting issue.
\looseness=-1

Fig.~\ref{fig:conv1x1} provides an example with size $1 \times 1$ kernels, which are typically applied later in the network to reduce the number of channels.
Therefore, they are usually quite deep and mapping single kernels to a full column of PEs across groups provides a way of reducing the worst-case memory footprint on the PEs, compared with mapping to a column in a single PE group. 
The accumulators are then set to forward the partial sums upwards to further accumulate with the partials sums generated by the PE groups above.
The same mapping is used for FC layers, where the rows of the weight matrix for an FC layer corresponds to the filters and the number of channels is equivalent to the width of the weight matrix.
\looseness=-1

\begin{table}[t!]
	\caption{Area for Lupulus synthesized for \SI{28}{\nano\metre} at \SI{1}{\giga\hertz}, and \SI{1}{\volt} (TT).}
	\label{tab:area}
		\centering
		\begin{tabular}{lrr}
			\toprule
			~ & \multicolumn{1}{c}{\SI{}{\micro\metre\squared} } & \multicolumn{1}{c}{Percentage} \\
			\midrule
			Accumulators					& \SI{387\,440}{}	& \SI{32.54}{\percent} \\
			Controllers						& \SI{3\,564}{} 	& \SI{0.30}{\percent} \\
			Input Buffers					& \SI{238\,110}{} 	& \SI{20.00}{\percent}  \\			
			Fetch Units						& \SI{9\,528}{} 	& \SI{0.80}{\percent} \\
			Networks						& \SI{43\,169}{}	& \SI{3.63}{\percent} \\
			PE Grid							& \SI{488\,325}{} 	& \SI{41.02}{\percent}  \\
			\midrule
			Total 							& \SI{1\,190\,600}{} 	& \\
			\bottomrule
	\end{tabular}
\end{table}

\section{Results}
\label{sec:results}

In this section, we first present the results of synthesizing Lupulus.
Then, we consider the performance of the HwA on a set of benchmarks and compare Lupulus with Eyeriss~\cite{Chen2016} and similar HwAs in terms of performance.

\subsection{Synthesis Results}
\label{ssec:area}

Table~\ref{tab:area} shows the area results for the HwA, synthesized using a \SI{28}{\nano\metre} FD-SOI technology with a \SI{1}{\volt} operating voltage and a target frequency of \SI{1}{\giga\hertz}. 
The input data is quantized to 8-bits and partial sums to 16-bits.
A grid of $15 \times 12$ PEs with $3 \times 3$ PE groups is used to match Eyeriss~\cite{Chen2016} in terms of maximum parallelism, with \SI{32}{\byte} for each PE, \SI{256}{\byte} for each input buffer, and \SI{2048}{\byte} for each output buffer, adding up to \SI{60.16}{\kilo\byte} of memory with double-buffering enabled for the input buffers and the SPMs in the PEs.
The PEs, accumulators, and input buffers use most of the area, with the control logic and networks of muxes taking up an insignificant portion.

\subsection{Benchmarks}
\label{ssec:benchmarks}

We now consider the result of benchmarking on the convolutional layers of AlexNet~\cite{krizhevsky2012imagenet} and \mbox{VGG-16}~\cite{simonyan2014very}.
While AlexNet is an older network, it contains layers with many different kernel sizes, which helps assess the limitations of the Lupulus architecture as it is optimized for the $3\times 3$ kernel size.
The benchmark results are generated using a high-level model of Lupulus. 
The model considers the time it takes to fetch, process and write out the results, given a memory interface with some bandwidth.
The hardware parameters are used to calculate the sizes of the different loops, similar to Alg.~\ref{alg:conv}, which determines the time for fetching/writing.
The computation time inside Lupulus is determined by the loop structure and the number of pipeline stages.

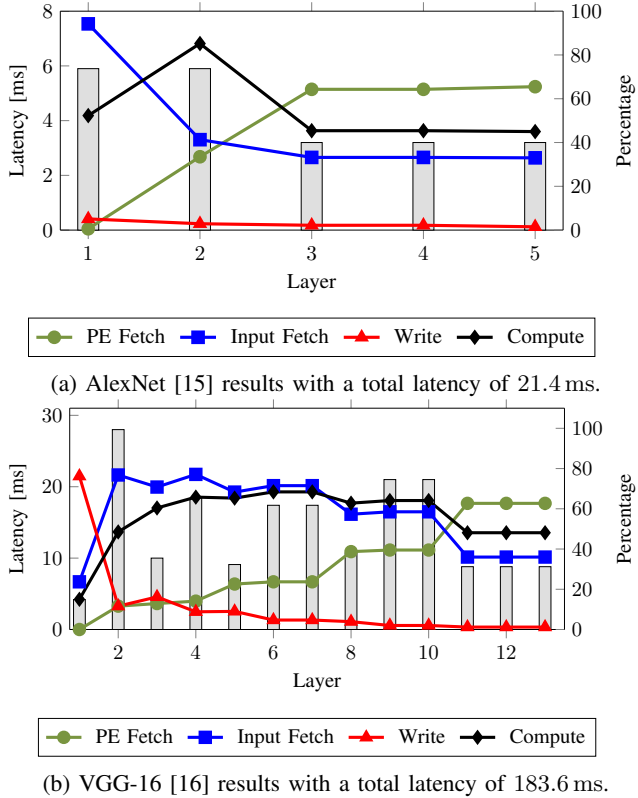
\begin{figure}[t]
	\centering
	\begin{subfigure}{\columnwidth}
		\centering
		\noindent\resizebox{\columnwidth}{!}{\begin{tikzpicture}
\pgfplotsset{
	xmin=0.75, xmax=5.25, width=10cm, height=5.25cm
}
% Latency
\begin{axis}[
ybar,
axis y line*=left,
ymin=0, ymax=8,
xlabel=Layer,
ylabel={Latency [ms]},
]
\addplot[black, fill=gray!25] table {data/alex_latency.dat};
%\addlegendentry{Latency}
\end{axis}

% Second axis, we have to add previous on top for legend
% Compute + I/O
\begin{axis}[
axis y line*=right,
axis x line=none,
ymin=0, ymax=100,
ylabel={Percentage},
legend columns = 4,
legend style = {at={(0.5, -0.40)}, anchor=north, inner sep=3pt, style={column sep=0.15cm}},       
legend cell align=left,
ytick = {0,20,40,60,80,100},
]
\addplot[mark=*,vheavygreen, line width=1.5pt,mark size=2.5pt] table {data/alex_pe_fetch.dat}; \addlegendentry{PE Fetch}
\addplot[mark=square*,blue, line width=1.5pt,mark size=2.5pt] table {data/alex_input_fetch.dat}; \addlegendentry{Input Fetch}
\addplot[mark=triangle*,red, line width=1.5pt,mark size=2.5pt] table {data/alex_write.dat}; \addlegendentry{Write}
\addplot[mark=diamond*,black, line width=1.5pt,mark size=2.5pt] table {data/alex_compute.dat}; \addlegendentry{Compute}
%			\legend{}; % empty the legend so as not to print it
\end{axis}
\end{tikzpicture}}
		\caption{AlexNet~\cite{krizhevsky2012imagenet} results with a total latency of \SI{21.4}{\milli\second}.}
		\label{fig:alexnet}
	\end{subfigure}
	\begin{subfigure}{\columnwidth}
		\centering
		\noindent\resizebox{\columnwidth}{!}{\begin{tikzpicture}
\pgfplotsset{
	xmin=0.75, xmax=13.5, width=10cm, height=5.35cm	
}
% Latency
\begin{axis}[
ybar,
bar width = 0.2cm,
axis y line*=left,
ymin=0, ymax=31,
xlabel=Layer,
ylabel={Latency [ms]},
]
\addplot[black, fill=gray!25] table {data/vgg_latency.dat};
\end{axis}

% Second axis, we have to add previous on top for legend
% Compute + I/O
\begin{axis}[
axis y line*=right,
axis x line=none,
ymin=0, ymax=110,
ylabel={Percentage},
ytick = {0,20,40,60,80,100},
legend columns = 4,
legend style = {at={(0.5, -0.40)}, anchor=north, inner sep=3pt, style={column sep=0.15cm}},       
legend cell align=left,     
]
\addplot[mark=*,vheavygreen, line width=1.5pt,mark size=2.5pt] table {data/vgg_pe_fetch.dat};
\addlegendentry{PE Fetch}
\addplot[mark=square*,blue, line width=1.5pt,mark size=2.5pt] table {data/vgg_input_fetch.dat};
\addlegendentry{Input Fetch}
\addplot[mark=triangle*,red, line width=1.5pt,mark size=2.5pt] table {data/vgg_write.dat};
\addlegendentry{Write}
\addplot[mark=diamond*,black, line width=1.5pt,mark size=2.5pt] table {data/vgg_compute.dat};
\addlegendentry{Compute}
%\legend{}
\end{axis}
\end{tikzpicture}}
		\caption{\mbox{VGG-16}~\cite{simonyan2014very} results with a total latency of \SI{183.6}{\milli\second}.}
		\label{fig:vgg16}
	\end{subfigure}%
	\caption{Per-layer latency and percentage of the total clock-cycles for different operations shown as bars and lines, respectively, for convolutional layers of AlexNet and \mbox{VGG-16}.}
	\label{fig:benchmarks}
\end{figure}

Fig.~\ref{fig:benchmarks} shows the latency of Lupulus for AlexNet and \mbox{VGG-16}.
The bars show the latency for each layer and the lines the percentage of clock cycles the HwA spends on fetching data to the PEs, fetching data to the input buffers, writing the results out, and computing the results. 
For the model, we assume that the external memory is connected through a 32-bit interface running at \SI{250}{\mega\hertz}, providing a bandwidth of $1\,\mathrm{GB/s}$. 
The on-chip frequency is set to the same as for synthesis, \SI{1}{\giga\hertz}.

For both networks, the memory interface is used almost \SI{100}{\percent} of the time, as shown in Fig.~\ref{fig:benchmarks}.
Initially, when the filters are shallow, the memory interface is dominated by fetching input feature map values.
Deeper in the network, the weight fetching dominates, as shown for the last layers in AlexNet and \mbox{VGG-16}.
The total number of clock cycles where the PEs are active also varies significantly, with \SI{80}{\percent} in layer 2 of AlexNet and \SI{40}{\percent} in the remaining convolutional layers.
While the on-chip clock-frequency can be slightly lowered to increase the utilization, with only a small increase in the total latency, the utilization can never reach \SI{100}{\percent} with a single channel memory interface.
With only a single channel, it is not possible to simultaneously read and write to the external memory, meaning that when the memories in the accumulators are full, the PEs stall.

While performance comparisons with other HwAs are difficult due to different architectural optimizations, process technologies, and methods of reporting results, we can compare our benchmark results against those of Eyeriss~\cite{Chen2016} due to similarities in the architectures and their level of complexity.
Table~\ref{tab:eyeriss_compare} shows the results for AlexNet and \mbox{VGG-16} when model parameters such as on-chip clock-frequency and speed of the external memory interface are adjusted to match Eyeriss.
For AlexNet, our HwA is approximately 3$\times$ slower, but close to twice as fast for \mbox{VGG-16}.
The difference for AlexNet is due to Eyeriss having two layers of memory, a shared global memory and local memories in the PEs for the inputs, weights, and outputs.
This makes Eyeriss able to better utilize the input feature maps for larger kernel sizes, whereas Lupulus has lower input reuse in this case.
However, Lupulus processes the smaller filters very efficiently as the inputs are streamed in and processed directly, whereas Eyeriss first has to load these into the PE memories.
As \mbox{VGG-16} represents a more modern workload, with small kernel sizes, the results for the current version of the HwA are satisfactory.

\begin{table}[t]
	\caption{Comparison of Lupulus HwA with Eyeriss~\cite{Chen2016}.}
%	\small	
	\label{tab:eyeriss_compare}
	\vspace*{-0.25em}	
	\centering
	\begin{tabular}{lrrr}
		\toprule
		\multicolumn{1}{l}{Network} & \multicolumn{1}{l}{Eyeriss} 	& \multicolumn{1}{c}{Ours} & \multicolumn{1}{c}{Speedup} \\
		\midrule
		AlexNet			& \SI{28.8}{\milli\second} & \SI{91.6}{\milli\second} &	\SI{0.31}{} \\
		\mbox{VGG-16}	& \SI{1436.5}{\milli\second} & \SI{773.0}{\milli\second} & \SI{1.86}{}	 \\
		\bottomrule
	\end{tabular}
\end{table}

\begin{table}[t]
	\caption{Performance comparison with similar HwAs, indicating the peak performance at the maximum frequency.}
%	\small
	\label{tab:accel_compare}
	\vspace*{-0.25em}	
%	\resizebox{\columnwidth}{!}{%	
		\centering
		\begin{tabular}{lrrrr}
			\toprule
			~ & \cite{Chen2016} & \cite{du2017reconfigurable} & \cite{cavigelli2016origami} & \textbf{Lupulus} 	  \\
			\midrule
			Process [nm] 				& $65$ 		& $65$ 		& $65$ 		& $28$ \\		
			Clk [GHz] 					& $0.25$ 	& $0.5$ 	& $0.75$ 	& $1.0$ \\													
			Gate Count [kGE] 			& $1176$ 	& $1300$ 	& $697$ 	& $799$ \\														
			Memory [kB] 				& $181.5$ 	& $112$ 	& $43$ 		& $60.61$ \\							
			Bit-width 					& $16$ 		& $16$ 		& $8$ 		& $8$ \\																								
			GOPS 						& $84$ 		& $152$ 	& $274$ 	& $380$ \\														
			GOPS/GHz 					& $336$ 	& $304$ 	& $365$ 	& $380$ \\																	
			\bottomrule
	\end{tabular}
\end{table}

Finally, Table~\ref{tab:accel_compare} compares similar accelerators, showing the process, the maximum frequency, the logic area, the amount of on-chip memory, the bit-width for inputs, and the GOPS (for reported frequency and normalized to \SI{1}{\giga\hertz}).
In terms of gate count and performance, Lupulus achieves the highest peak performance among the considered accelerators, at a \SI{32}{\percent} and \SI{39}{\percent} lower gate count compared to~\cite{Chen2016} and~\cite{du2017reconfigurable}, respectively.
However, we use \SI{14}{\percent} more gates than~\cite{cavigelli2016origami} with only a \SI{4}{\percent} improvement in GOPS/GHz.

\section{Conclusion}
\label{sec:conclusion}

In this paper, we described Lupulus, a flexible hardware architecture supporting different types of NN architectures.
Lupulus provides the capability of merging different groups of PEs or overlap small convolutional kernels inside groups to improve the utilization of the local memories and the PEs depending on the type of network.
Lupulus can be optimized for a specific kernel size to maximize the performance, while still supporting the execution of different NNs efficiently, with just a single layer of memory.
Lupulus was implemented in a \SI{28}{\nano\meter} FD-SOI technology utilizing \SI{60}{\kilo\byte} of on-chip memory and demonstrating a peak performance of $380\,\mathrm{GOPS/GHz}$ with latencies of \SI{21.4}{\milli\second} and \SI{183.6}{\milli\second} for the convolutional layers of AlexNet and \mbox{VGG-16}, respectively.

%\clearpage

\balance
\bibliographystyle{IEEEtran}
\bibliography{refs}

% Generated by IEEEtran.bst, version: 1.14 (2015/08/26)
\begin{thebibliography}{10}
\providecommand{\url}[1]{#1}
\csname url@samestyle\endcsname
\providecommand{\newblock}{\relax}
\providecommand{\bibinfo}[2]{#2}
\providecommand{\BIBentrySTDinterwordspacing}{\spaceskip=0pt\relax}
\providecommand{\BIBentryALTinterwordstretchfactor}{4}
\providecommand{\BIBentryALTinterwordspacing}{\spaceskip=\fontdimen2\font plus
\BIBentryALTinterwordstretchfactor\fontdimen3\font minus
  \fontdimen4\font\relax}
\providecommand{\BIBforeignlanguage}[2]{{%
\expandafter\ifx\csname l@#1\endcsname\relax
\typeout{** WARNING: IEEEtran.bst: No hyphenation pattern has been}%
\typeout{** loaded for the language `#1'. Using the pattern for}%
\typeout{** the default language instead.}%
\else
\language=\csname l@#1\endcsname
\fi
#2}}
\providecommand{\BIBdecl}{\relax}
\BIBdecl

\bibitem{hinton2012deep}
G.~{Hinton}, L.~{Deng}, D.~{Yu}, G.~E. {Dahl}, A.~{Mohamed}, N.~{Jaitly},
  A.~{Senior}, V.~{Vanhoucke}, P.~{Nguyen}, T.~N. {Sainath}, and
  B.~{Kingsbury}, ``Deep neural networks for acoustic modeling in speech
  recognition: The shared views of four research groups,'' \emph{IEEE Signal
  Processing Magazine}, vol.~29, no.~6, pp. 82--97, Nov. 2012.

\bibitem{schmidhuber2015deep}
\BIBentryALTinterwordspacing
J.~Schmidhuber, ``Deep learning in neural networks: An overview,'' \emph{Neural
  Networks}, vol.~61, p. 85–117, Jan. 2015. [Online]. Available:
  \url{http://dx.doi.org/10.1016/j.neunet.2014.09.003}
\BIBentrySTDinterwordspacing

\bibitem{LeCun2015}
Y.~LeCun, Y.~Bengio, and G.~Hinton, ``Deep learning,'' \emph{Nature}, vol. 521,
  no. 7553, pp. 436--444, May 2015.

\bibitem{He2015}
K.~He, X.~Zhang, S.~Ren, and J.~Sun, ``Deep residual learning for image
  recognition,'' in \emph{IEEE Conference on Computer Vision and Pattern
  Recognition (CVPR)}, June 2016.

\bibitem{Chen2016}
Y.~{Chen}, T.~{Krishna}, J.~S. {Emer}, and V.~{Sze}, ``Eyeriss: {A}n
  energy-efficient reconfigurable accelerator for deep convolutional neural
  networks,'' \emph{IEEE Journal of Solid-State Circuits}, vol.~52, no.~1, pp.
  127--138, Jan. 2017.

\bibitem{han2016eie}
\BIBentryALTinterwordspacing
S.~Han, X.~Liu, H.~Mao, J.~Pu, A.~Pedram, M.~A. Horowitz, and W.~J. Dally,
  ``{EIE}: Efficient inference engine on compressed deep neural network,'' in
  \emph{Annual International Symposium on Computer Architecture (ISCA)}, 2016,
  pp. 243--254. [Online]. Available: \url{https://doi.org/10.1109/ISCA.2016.30}
\BIBentrySTDinterwordspacing

\bibitem{wei2017automated}
X.~Wei, C.~H. Yu, P.~Zhang, Y.~Chen, Y.~Wang, H.~Hu, Y.~Liang, and J.~Cong,
  ``Automated systolic array architecture synthesis for high throughput {CNN}
  inference on {FPGAs},'' in \emph{ACM/EDAC/IEEE Design Automation Conference
  (DAC)}, June 2017, pp. 1--6.

\bibitem{du2017reconfigurable}
L.~{Du}, Y.~{Du}, Y.~{Li}, J.~{Su}, Y.~{Kuan}, C.~{Liu}, and M.~F. {Chang}, ``A
  reconfigurable streaming deep convolutional neural network accelerator for
  internet of things,'' \emph{IEEE Transactions on Circuits and Systems I:
  Regular Papers}, vol.~65, no.~1, pp. 198--208, Jan. 2018.

\bibitem{andri2018yodann}
R.~{Andri}, L.~{Cavigelli}, D.~{Rossi}, and L.~{Benini}, ``{YodaNN}: An
  architecture for ultralow power binary-weight cnn acceleration,'' \emph{IEEE
  Transactions on Computer-Aided Design of Integrated Circuits and Systems},
  vol.~37, no.~1, pp. 48--60, Jan 2018.

\bibitem{hegde2018ucnn}
K.~Hegde, J.~Yu, R.~Agrawal, M.~Yan, M.~Pellauer, and C.~W. Fletcher, ``{UCNN}:
  Exploiting computational reuse in deep neural networks via weight
  repetition,'' in \emph{Annual International Symposium on Computer
  Architecture (ISCA)}, 2018, pp. 674--687.

\bibitem{yin2017high}
S.~Yin, P.~Ouyang, S.~Tang, F.~Tu, X.~Li, S.~Zheng, T.~Lu, J.~Gu, L.~Liu, and
  S.~Wei, ``A high energy efficient reconfigurable hybrid neural network
  processor for deep learning applications,'' \emph{IEEE Journal of Solid-State
  Circuits}, vol.~53, no.~4, pp. 968--982, 2017.

\bibitem{desoli2017}
G.~Desoli, N.~Chawla, T.~Boesch, S.-p. Singh, E.~Guidetti, F.~De~Ambroggi,
  T.~Majo, P.~Zambotti, M.~Ayodhyawasi, H.~Singh, and N.~Aggarwal, ``A 2.9
  {TOPS/W} deep convolutional neural network {SoC} in {FS-SOI} 28nm for
  intelligent embedded systems,'' in \emph{IEEE International Solid-State
  Circuits Conference (ISSCC)}, 2017, pp. 238--239.

\bibitem{luo2016dadiannao}
T.~Luo, S.~Liu, L.~Li, Y.~Wang, S.~Zhang, T.~Chen, Z.~Xu, O.~Temam, and
  Y.~Chen, ``Da{D}ian{N}ao: {A} neural network supercomputer,'' \emph{IEEE
  Transactions on Computers}, vol.~66, no.~1, pp. 73--88, Jan. 2016.

\bibitem{lecun2010convolutional}
Y.~{LeCun}, K.~{Kavukcuoglu}, and C.~{Farabet}, ``Convolutional networks and
  applications in vision,'' in \emph{IEEE International Symposium on Circuits
  and Systems (ISCAS)}, May 2010, pp. 253--256.

\bibitem{krizhevsky2012imagenet}
\BIBentryALTinterwordspacing
A.~Krizhevsky, I.~Sutskever, and G.~E. Hinton, ``Image{N}et classification with
  deep convolutional neural networks,'' \emph{Communications of the ACM},
  vol.~60, no.~6, pp. 84--90, May 2017. [Online]. Available:
  \url{http://doi.acm.org/10.1145/3065386}
\BIBentrySTDinterwordspacing

\bibitem{simonyan2014very}
K.~Simonyan and A.~Zisserman, ``Very deep convolutional networks for
  large-scale image recognition,'' \emph{arXiv 1409.1556}, Sep. 2014.

\bibitem{szegedy2015going}
C.~Szegedy, W.~Liu, Y.~Jia, P.~Sermanet, S.~Reed, D.~Anguelov, D.~Erhan,
  V.~Vanhoucke, and A.~Rabinovich, ``Going deeper with convolutions,'' in
  \emph{IEEE Conference on Computer Vision and Pattern Recognition (CVPR)},
  2015, pp. 1--9.

\bibitem{Sze2017}
V.~{Sze}, Y.~{Chen}, T.~{Yang}, and J.~S. {Emer}, ``Efficient processing of
  deep neural networks: {A} tutorial and survey,'' \emph{Proceedings of the
  IEEE}, vol. 105, no.~12, pp. 2295--2329, Dec 2017.

\bibitem{jouppi2017datacenter}
N.~P. Jouppi, C.~Young, N.~Patil, D.~Patterson, G.~Agrawal, R.~Bajwa, S.~Bates,
  S.~Bhatia, N.~Boden, A.~Borchers \emph{et~al.}, ``In-datacenter performance
  analysis of a tensor processing unit,'' in \emph{Annual International
  Symposium on Computer Architecture (ISCA)}, 2017, pp. 1--12.

\bibitem{cavigelli2016origami}
L.~Cavigelli and L.~Benini, ``{Origami}: {A} 803-gop/s/w convolutional network
  accelerator,'' \emph{IEEE Transactions on Circuits and Systems for Video
  Technology}, vol.~27, no.~11, pp. 2461--2475, 2016.

\end{thebibliography}

\end{document}